\documentclass[conference]{IEEEtran}
\IEEEoverridecommandlockouts

\usepackage{float}
\usepackage{amsmath}
\usepackage{caption}
\usepackage{subcaption}
\usepackage{hyperref}

\usepackage{cite}

\ifCLASSINFOpdf
   \usepackage[pdftex]{graphicx}
\else
\fi
\begin{document}
%
\title{Delineating Intra-Urban Spatial Connectivity Patterns by Travel-Activities: A Case Study of Beijing, China}

\author{\IEEEauthorblockN{Chaogui Kang\IEEEauthorrefmark{2}}
\IEEEauthorblockA{Institute of Remote Sensing and \\Geographical Information Systems\\
Peking University\\
Beijing, P.R. China 100871\\
Email: chaoguikang@pku.edu.cn}
\and
\IEEEauthorblockN{Yu Liu}
\IEEEauthorblockA{Institute of Remote Sensing and \\Geographical Information Systems\\
Peking University\\
Beijing, P.R. China 100871\\
Email: liuyu@urban.pku.edu.cn}
\and
\IEEEauthorblockN{Lun Wu}
\IEEEauthorblockA{Institute of Remote Sensing and \\Geographical Information Systems\\
Peking University\\
Beijing, P.R. China 100871\\
Email: wulun@pku.edu.cn}}

\maketitle

\begin{abstract}
Travel activities have been widely applied to quantify spatial interactions between places, regions and nations. In this paper, we model the spatial connectivities between 652 Traffic Analysis Zones (TAZs) in Beijing by a taxi OD dataset. First, we unveil the gravitational structure of intra-urban spatial connectivities of Beijing. On overall, the inter-TAZ interactions are well governed by the Gravity Model $G_{ij} = \lambda p_{i}p_{j} / d_{ij}$, where $p_{i}$, $p_{j}$ are degrees of TAZ $i$, $j$ and $d_{ij}$ the distance between them, with a goodness-of-fit around $0.8$. Second, the network based analysis well reveals the polycentric form of Beijing. Last, we detect the semantics of inter-TAZ connectivities based on their spatiotemporal patterns. We further find that inter-TAZ connections deviating from the Gravity Model can be well explained by link semantics. 
\end{abstract}



\section{Introduction}
With the explosion of geospatial data in past few years, data-driven urban analysis has been emerging for deep understanding the urban environment. Topics including human mobility\cite{Liu:2012a}, urban configuration\cite{Liu:2012b}, transport intelligence\cite{Yuan:2010}, energy and pollution\cite{Zheng:2013} benefit substantially from the so-called ``Big Data Revolution"\cite{Mayer-Schönberger:2013}. Trajectory data with detailed spatiotemporal information is of particular interests to geographers\cite{Lu:2012}. This kind of data provides a promising tool for exploring the interplay between human travel activities and the built urban environment. By aggregating individuals' movements, the spatial interactions between different sub-zones within a city can be easily obtained\cite{Kang:2013}.

Recently, many structural properties of the intra-urban spatial interaction network have been explored based on human travel activities\cite{Zhong:2014}. In this research, we investigate the spatiotemporal characteristics of taxi flows between 652 Traffic Analysis Zones (TAZs) in Beijing. The objective includes: (1) modeling the global structure of the inter-TAZ taxi interaction network. More specifically, we testify whether the network follows the Gravity Model. It will help us to understand how TAZs are interconnected with each other; (2) detecting the major centers within the study area. It can uncover the functional (sub-)regions and the sources and sinks of human travel-activities in the study area;  (3) characterizing the temporal fluctuations of inter-TAZ flows. It can differentiate travel flows associated with different purposes and activities. Putting together, these three aspects will give a detailed depiction of the intra-urban taxi interaction network.

The remainder of this article is organized as follows. Section II describes the taxi trajectories we utilized to extract the Origin-Destination (OD) flows between different TAZs. Section III introduces the methodology we adopted to quantify the structural properties of the inter-TAZ network. Approaches for constructing the spatial interaction network, reverse-fitting the gravity model, and detecting the semantics of inter-TAZ flows are given in this section. Then, Section IV presents our findings of the gravitational and the polycentric structures of the inter-TAZ network as well as the semantics of the inter-TAZ flows. Finally, Concluding remarks are given in Section V.


\section{Data}
In this research, we leverage a dataset of taxi GPS trajectories collected between November 1, 2012 and November 30, 2012 in Beijing, China for analysis. The entire dataset covers 12,000 taxicabs, and contains detailed information of taxi-ID, passenger pick-up location, passenger pickup time, passenger drop-off location, passenger drop-off time and path travelled of each taxi trip. The typical time gap between two consecutive GPS points is 10 seconds or one minute. On a daily basis, more than 30 million GPS points are collected, capturing about 0.3 million taxi trips (or OD pairs) within the study area\footnote{Note that we adopt the traffic analysis zones within the fifth ring road of Beijing as the study area and take each TAZ as a basic analysis unit.}.

Temporally, the usage of taxicabs within the study area shows significant rhythms. In general, there are more passengers on board during the commuting periods than non-commuting periods, resulting in a ``bi-modal" or ``tri-modal" distribution of the number of taxi trips captured in each hour. Furthermore, there are more pick-up points than drop-off points within the study area in the morning and the noon, implying a lot of people are traveling in these time slots. Besides, there are also more taxi trips in weekdays than weekends in the study area.

Spatially, taxi trips highly concentrate at major commercial centers and transport hubs in the study area. Strong positive spatial autocorrelations are observed for both the distribution of trip origins and the distribution of trip destinations. The Local Indicators of Spatial Autocorrelation (LISA\cite{Anselin:1995}) of taxi ODs within TAZs\footnote{To calculate the LISA index, we count the taxi origins and destinations within each TAZ. The high LISA value demonstrates the spatial heterogeneity of taxi OD distribution.} are close to $0.55$. The distributions of the number of taxi ODs in each TAZ generally follow the exponential distribution. Additionally, the number of taxi pick-ups and the number of taxi drop-offs in a TAZ are highly balanced in a day, with a Pearson Correlation Coefficient (PCC) of 0.945 for weekdays and 0.946 for weekends.


\section{Methods}
\subsection{Network Construction}
We allocate the pickup point and the drop-off point of each taxi trip into TAZs, and then aggregate the number of trips between every pair of TAZs (during the 22 weekdays and the 8 weekends respectively). To construct a spatial-embedded network, we first take each of the 652 TAZs as a node $i$ and the coordinates $(x_{i}, y_{i})$ of its centroid as the spatial location of the node. Second, we assign a directed edge $E_{i, j}$ to a pair of nodes $(i, j)$ if there are taxi trips connecting their corresponding TAZs. Third, the assigned link  $E_{i, j}$ is weighted by the number of taxi trips departing from the starting node $i$ and arriving at the ending node $j$. By doing so, we obtain a $652 \times 652$ network representing the spatial interactions between different sub-areas (TAZs) within the study area.

For weekdays, the resulting network \textbf{$N_{1}$} has $348,065$ edges, an average (weighted) node degree ${\langle k_{1} \rangle}$ of $45,125$ and a clustering coefficient $\langle c_{1} \rangle$ of $0.686$. Whereas, the resulting network \textbf{$N_{2}$} for weekends has $279,409$ edges, an average node degree ${\langle k_{2} \rangle}$ of $13,548$ and a clustering coefficient $\langle c_{2} \rangle$ of $0.535$.

\subsection{Gravity Model Fitting}
The general form of the Gravity Model can be written as
\begin{align} 
   G_{ij} = \lambda \frac{P_{i} P_{j}}{d_{ij}^{\beta}}
   \label{eq:gm}
\end{align}
where $G_{ij}$ denotes the flow (or interaction) between two locations $(i, j)$ with a size (or attraction) of $P_{i}$ and $P_{j}$ respectively; $d_{ij}$ quantifies the deterrence (in terms of distance or time) between $i$ and $j$; $\beta$ is the distance-decay parameter and $\lambda$ a scaling constant. This highly simplified model has been criticized for a couple of limitations, including symmetric structure and non-rigorous derivation\cite{Simini:2012}. However, the beauty of a clear form and the capability of revealing global (or macro-scale) interaction patterns make the Gravity Model still attractive to geographers. In this paper, we fit this simple model to network $N_{1}$ and $N_{2}$ and obtain substantial consistency between the model and the observed inter-TAZ interaction patterns. For simplicity, we use the general term ``Gravity Model" to represent model of the form of Equation \ref{eq:gm} specifically hereafter.

In reality, we usually have explicit observations of the spatial interaction $G_{ij}$ and the distance $d_{ij}$ between different locations. Also, since $\lambda$ is a scaling constant and plays marginal role in the interacting system, its value are usually pre-defined in practice. In this sense, the objective of fitting the Gravity Model is to estimate the parameters $P_{i}$, $P_{j}$ and $\beta$ in Equation \ref{eq:gm}. Two distinct techniques have been extensively applied to address this problem in existing literature. The first approach is linear programming \cite{Okelly:1995}, in which the Gravity Model is transformed into a linear system as
\begin{align} 
   ln{P_{i}} + ln{P_{j}} - (ln{d_{ij}}) \beta = ln{G_{ij}} - ln{\lambda}
   \label{eq:linear}
\end{align}
This linear programming system can be solved by ordinary MINIMAX, MAD and GP methods\footnote{Pease refer to \cite{Okelly:1995} for detailed information about MINIMAX, MAD and GP methods.}. The second approach is the algebraic method \cite{Shen:2004}, which is an approximation algorithm by conducting series multiplication of Equation \ref{eq:gm} as 
\begin{align*} 
   \prod_{i \neq j}{G_{ij}} = \prod_{i \neq j}{\lambda P_{i} P_{j} d^{-\beta}} \text{ or } \prod_{j = i+1}{G_{ij}} = \prod_{j = i+1} {\lambda P_{i} P_{j} d^{-\beta}}
   \label{eq:multiple}
\end{align*}
This multiplication then gives the relation 
\begin{align}
\prod_{i=1}^{n} P_{i} & = \sqrt[n-1]{\prod_{i=1, j=i+1}^{n} \lambda G_{ij}d_{ij}^{-\beta}} 
\end{align}
Similarly, do the multiplication for the links from $P_{i}$ to all other nodes and obtain
\begin{align}
P_{i}^{n-2}\prod_{j=1, j \neq i}^{n} P_{j} & = \prod_{j=1, j \neq i}^{n}{P_{i}P_{j}} & =  \lambda^{n} \prod_{j=1, j \neq i}^{n} G_{ij}d_{ij}^{-\beta}
\end{align}
Then, $P_{i}$ is calibrated as a function of $\beta$ by dividing Equation 4 by Equation 3. Finally, $\beta$ is tweaked from $0$ to $2$ with an increment $\delta$ (usually equal to $0.1$) for predicting flows $\hat{G_{ij}}$ between different locations. Eventually, $\beta$ and $P_{i}$ producing results with the highest consistency to observed flows are taken as the model parameters.  

Unfortunately, the linear programing approach has a remarkably high time complexity and is incapable to process a network with 652 nodes on our computation environment. Therefore, we adopt the algebraic method in this research and our data and codes can be downloaded at \url{http://pkugeosoft.org/Resource.aspx}.

\subsection{Sub-Center Identification}
To find functional (sub-)regions in the study area, we conduct the community detection on the inter-TAZ network. A thorough introduction of community detection in graphs is given in \cite{Fortunato:2010}. In this article, a functional (sub-)region is treated as a cluster of TAZs within which intensive interactions exist. By mapping the links with a large weight on space, this method assists us to easily identify sub-centers within the study area. 

\subsection{Link Semantic Detection}
For each pair of TAZs, we build a link signature capturing temporal fluctuations of the number of taxi trips between them during a day. Taking one-hour as the temporal granularity, the signature $S_{ij}$ of a directed link from $i$ to $j$ is denoted as a $1 \times 24$ vector 
\begin{align} 
   S_{ij} = [T_{ij}^{1}, T_{ij}^{2}, \cdots, T_{ij}^{23}, T_{ij}^{24}]
\end{align}
where $T_{ij}^{t}$ is the number of taxi trips from $i$ to $j$ in a given time slot $t$. Linking with the activity chain of daily movements, this signature contains rich context information and can be used to uncover the social and functional properties of its origin and destination as well as the link itself (termed as ``link semantics"). For instance, the daily routine of individuals' travel activities typically follows the pattern ``home $\rightarrow$ workplace $\rightarrow$ restaurant (or home) $\rightarrow$ workplace $\rightarrow$ home". For the commuting flow between a residential (sub-)area $i$ and a commercial (sub-)area $j$, heavy taxi traffics will be observed from $i$ to $j$ between $7:00$ and $9:00$, and from $j$ to $i$ between $17:00$ and $19:00$. 

In this article, we detect typical temporal interaction patterns by clustering the link signatures of pairs of TAZs. To minimize the influence of casual taxi usages, we only analyze links with more than $125$ taxi trips averaged by each day\footnote{We choose the threshold 125 due to that it guarantees a reasonable number of trips in each hour and total valid links remain for following analysis.}. In the weekday network $N_{1}$, 748 valid links (with a weight $w \geq 125 \times 22 =2750$) are extracted. Similarly, 742 valid links (with a weight $w \geq 125 \times 8 = 1000$) are obtained from the weekend network $N_{2}$. As discussed in Section II, taxi trips are heterogeneously distributed in space. We thus normalize the link signatures and the normalized signature $S_{ij}^{norm}$ is built based on the z-score principle as
\begin{align} 
   S_{ij}^{norm} & = [\frac{T_{ij}^{1} - \mu}{\sigma}, \frac{T_{ij}^{2} - \mu}{\sigma}, \cdots, \frac{T_{ij}^{23} - \mu}{\sigma}, \frac{T_{ij}^{24} - \mu}    {\sigma}] \nonumber \\ & = [Z_{ij}^{1}, Z_{ij}^2, \cdots, Z_{ij}^{23}, Z_{ij}^{24}]
\end{align}
where $\mu$ is the mean of $T_{ij}^{t}$ on 24 hours and $\sigma$ is the standard deviation. Considering the global pattern of $S_{ij}^{norm}$ is largely determined by the intensity of human activities, we further transform $S_{ij}^{norm}$ to $S_{ij}^{res}$ as
\begin{align} 
   S_{ij}^{res} & = [Z_{ij}^{1} - z^{1},Z_{ij}^{2} - z^{2}, \cdots, Z_{ij}^{23} - z^{23}, Z_{ij}^{24} - z^{24}] \nonumber \\ & = [R_{ij}^{1}, R_{ij}^2, \cdots, R_{ij}^{23}, R_{ij}^{24}]
   \end{align}
where $z^{t}$ is the average of $Z_{ij}^{t}$ on all valid links at a given time $t$. The transformation produces better visualization of the differences between link signatures. It is also notable that clustering results of $S_{ij}^{norm}$  and $S_{ij}^{res}$ are identical in that the subtraction of $z^{t}$ and $Z_{ij}^{t}$ results in no differences of the similarity between signatures.

	The X-means algorithm (provided in the WEKA package \cite{Hall:2009}) is utilized to identify typical interaction signatures. This algorithm automatically optimizes the number of clusters based on the Bayesian Information Criterion (BIC) or the Akaike Information Criterion (AIC) principles\cite{Pelleg:2000}. Since the clustering of link signatures is an unsupervised process, X-means is an adequate tool for our analysis. In this research, four typical clusters of signatures are detected and mapped on space to illustrate the semantics of the links. With prior knowledge of the spatial distributions of the functional areas within the study area, these typical signatures also depict the interplay between different functional areas of the city.

	 
\section{Results}
\subsection{Gravitational Structure}

\begin{figure}[htbp]
        \centering
        \begin{subfigure}[b]{0.315\textwidth}
                \includegraphics[width=\textwidth]{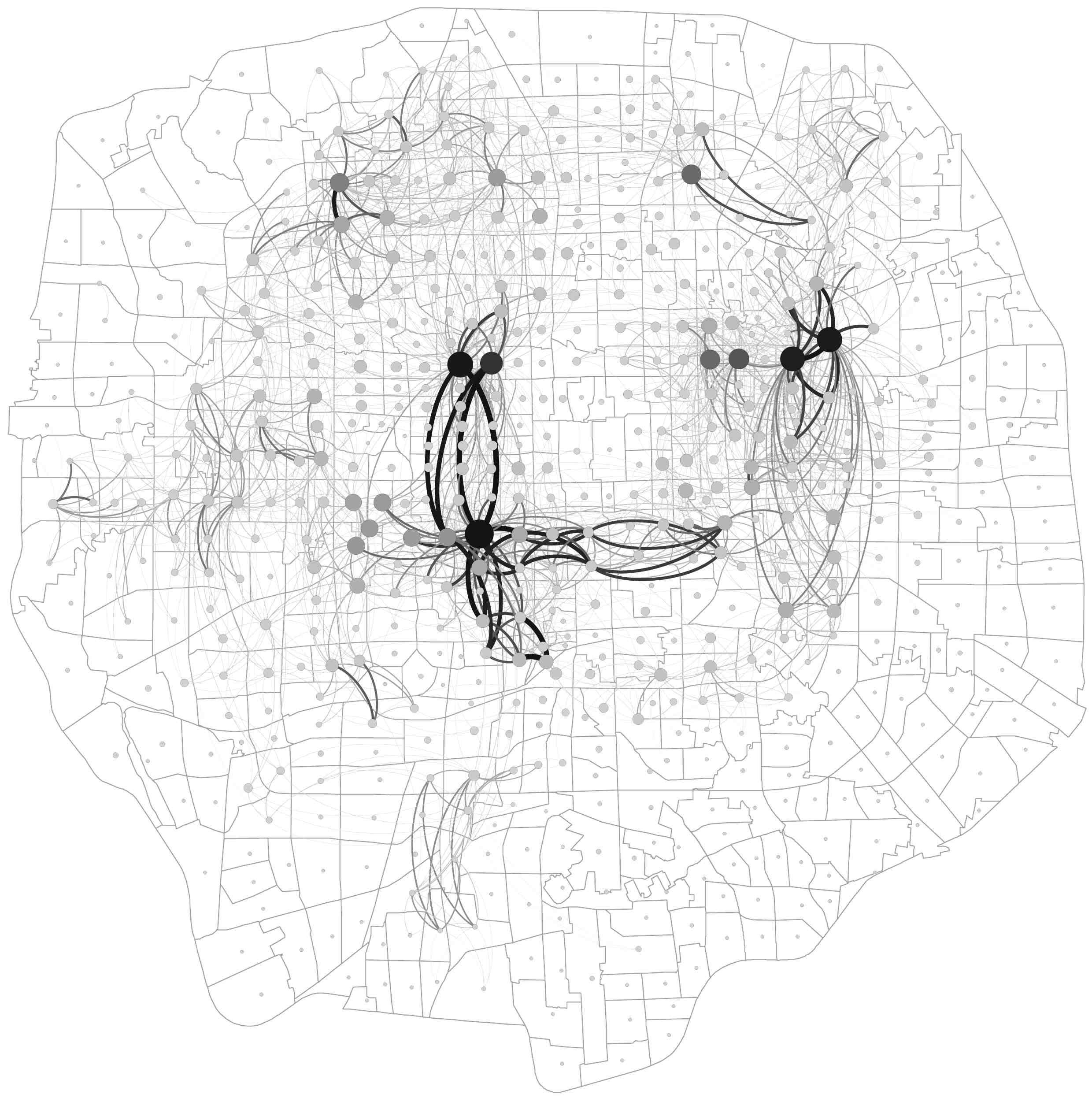}
                \caption{Observation}
                \label{fig:empirical}
        \end{subfigure}%
        ~\\ 
        \begin{subfigure}[b]{0.315\textwidth}
                \includegraphics[width=\textwidth]{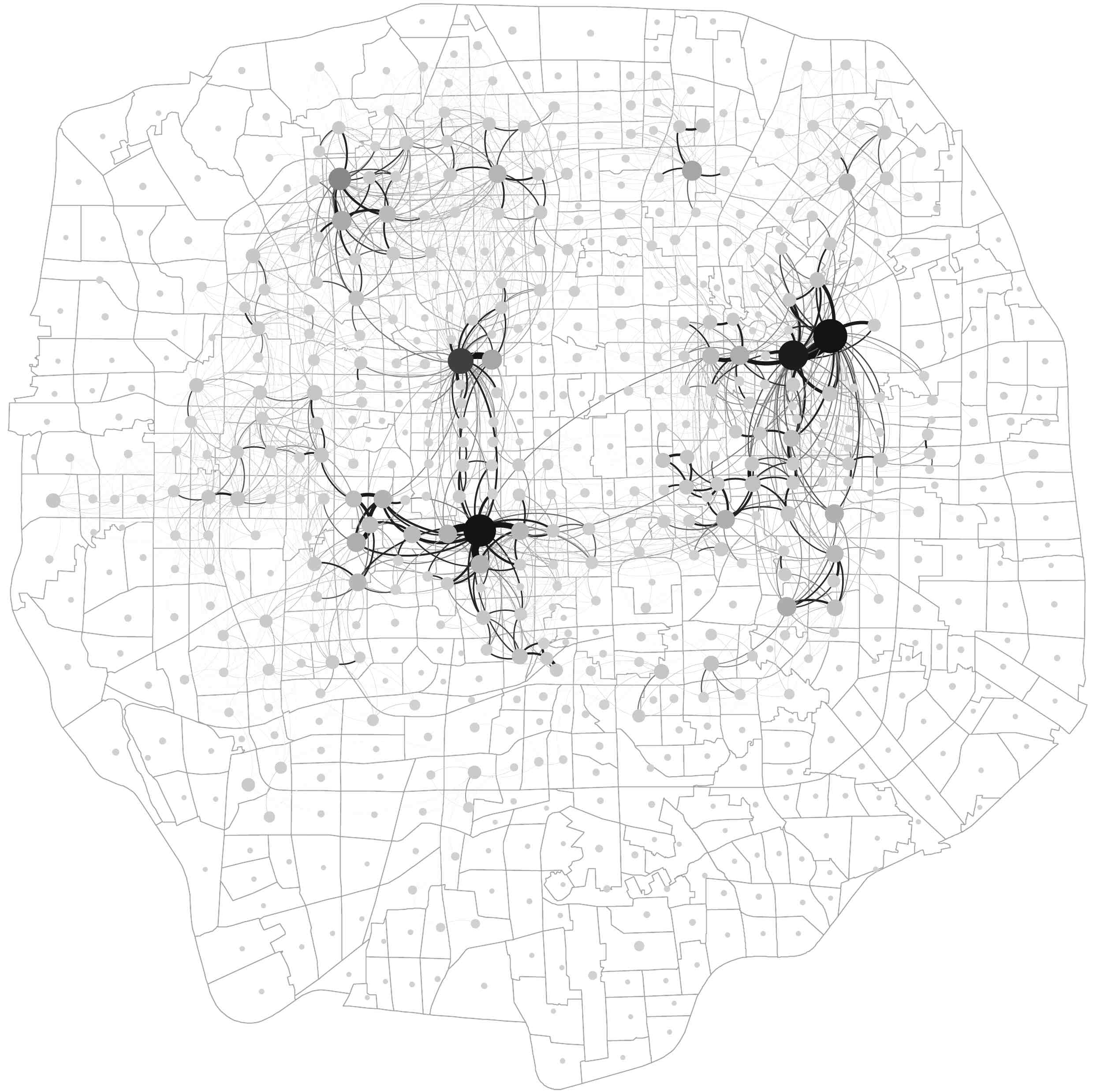}
                \caption{Gravity Model}
                \label{fig:model}
        \end{subfigure}
        ~\\ 
        \begin{subfigure}[b]{0.315\textwidth}
                \includegraphics[width=\textwidth]{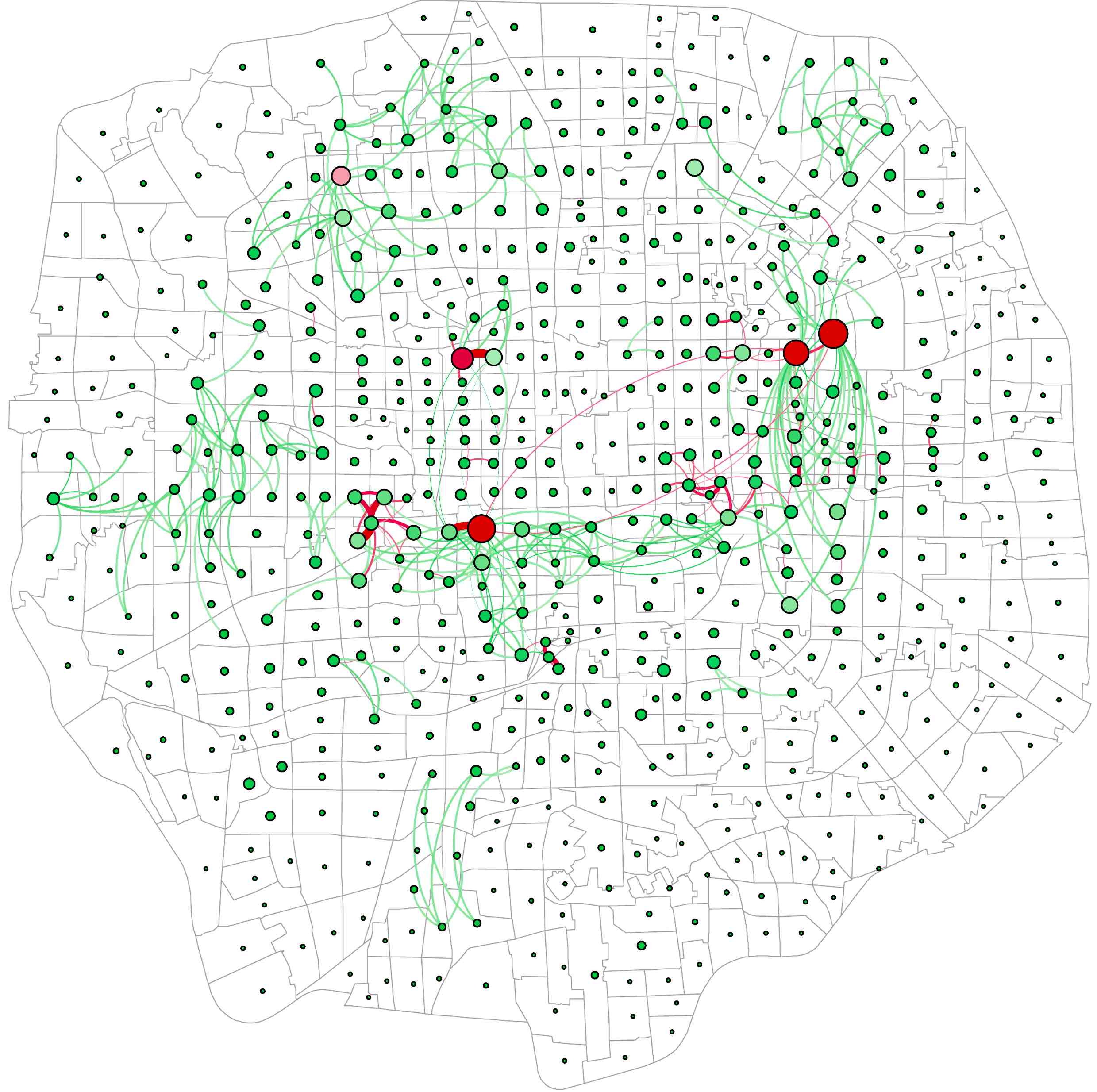}
                \caption{Residual}
                \label{fig:diff}
        \end{subfigure}
        \label{fig:gm}
        \caption{Spatial network of inter-TAZ taxi OD flows. (a) The observed inter-TAZ taxi flows; (b) The estimated inter-TAZ taxi flows by the Gravity Model; (c) The differences between the two networks. Weights of edges in green color are underestimated by the model, whereas weights of edges in red color are overestimated. The correlation between the observed network and the estimated network is 0.824, implying that the inter-TAZ network well follows the proposed Gravity Model. Note that the size of each node represents its degree or attraction.}\label{fig:network}
\end{figure}

The taxi OD network is well fitted by the Gravity Model with $\beta$ equal to 1. Figure 1(a) and 1(b) demonstrate the empirical network and the network estimated by the Gravity Model of weekdays, respectively. Note that the spatial structure of the weekend network is similar with the weekday network, and thus is excluded for simplicity. On overall, spatial patterns of the observed network and the modeled network are well matched. The Goodness-of-Fit (GOF)\footnote{In this article, the GOF is defined as the PCC (or R Square) between the observed inter-TAZ flows and the estimated inter-TAZ flows by the Gravity Model.} is $0.824$ for the weekday network $N_{1}$ and $0.783$ for the weekend network $N_{2}$ (see Figure 2). However, several mismatches are also identified as shown in Figure 1(c), within which those links in green color are underestimated and links in red color are overestimated by the proposed model. Here we argue that the Gravity Model $G_{ij} = \lambda P_{i} P_{j} / d_{ij}$ well governs the global organization of inter-TAZ interactions since the GOF is acceptable. We further investigate the mismatched parts between the observations and the model in following subsections.   

\begin{figure}[htbp]
        \centering
        \begin{subfigure}[b]{0.45\textwidth}
                \includegraphics[width=0.475\textwidth]{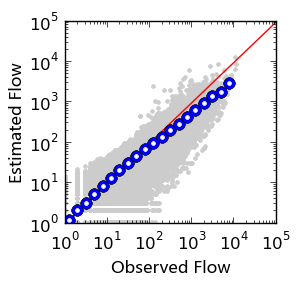}
                \includegraphics[width=0.475\textwidth]{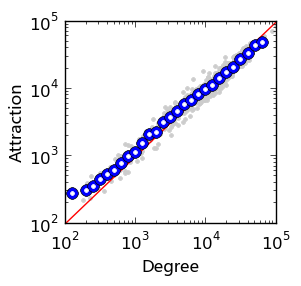}
                \caption{Weekday}
                \label{fig:weekday}
        \end{subfigure}%
        ~ \\
        \begin{subfigure}[b]{0.45\textwidth}
                \includegraphics[width=0.475\textwidth]{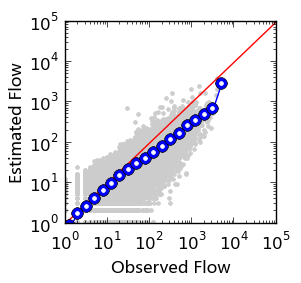}
                \includegraphics[width=0.475\textwidth]{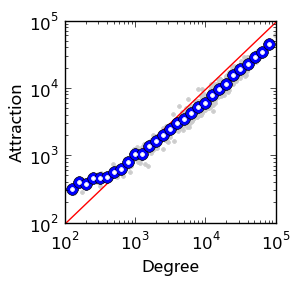}
                \caption{Weekend}
                \label{fig:weekend}
        \end{subfigure}
            \caption{Validation of the fitted Gravity Model. Plots on the left show the correlation between the observed inter-TAZ flows and the flows estimated by the Gravity Model. Plots on the right demonstrate the correlation between the node degrees and the estimated node attractions in the network. All the correlations are very high, indicating the proposed model is acceptable.}\label{fig:gof}
\end{figure}

The Gravity Model also estimates the attraction $P_{i}$ of each node (TAZ) in the network. We then explore the relationship between node attractions and node degrees to better understand inter-TAZ flows. The high correlations between the node degrees and the estimated node attractions as shown in Figure 2 imply that the node attraction $P_{i}$ can be simply replaced by the node degree $p_{i}$ in the context of this research. Quantitatively, the PCC of $P_{i}$ and $p_{i}$ is $0.976$ for weekdays and $0.977$ for weekends. In other words, the best model fitting with our inter-TAZ networks can be further simplified as 
\begin{align} 
   G_{ij} = \lambda \frac{p_{i} p_{j}} {d_{ij}}
\end{align}

\subsection{Polycentricity}
Under closer scrutiny, clusters of nodes with large attractions are identified in Figure 1, implying a polycentric form of the study area. Therefore, we map all the edges with a weight above the given threshold $1,000$ in Figure 3. In the plot, each node represents a TAZ, and the color of the node is assigned based on the community detection results of the entire network. In the context of this research, 14 distinct spatial cohesive communities are detected by the COMBO algorithm\cite{Sobolevsky:2013} with a modularity of $0.281$. Note that each community represents a cohesive cluster of TAZs with intensive interactions with each other in space. The links between pairs of nodes are then colored by the communities of its origin and destination. Obviously, links within a identical community dominates in the inter-TAZ network, which is consistent with the results of community detection. 

\begin{figure}[htbp]
   \centering
   \includegraphics[width = 0.37\textwidth]{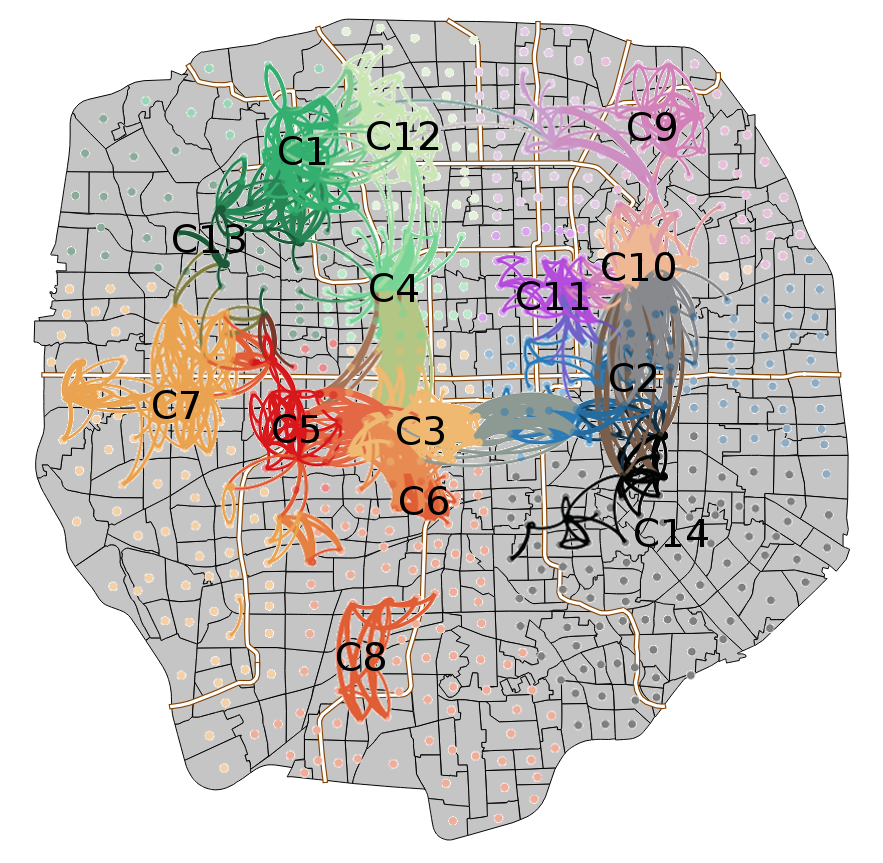} 
   \caption{Polycentric structure revealed by taxi OD network. Edges with a weight less than 1000 are filtered out in visualization. The study area is divided into 14 clusters by the community detection algorithm. Each node (or TAZ centroid) is assigned a color representing its assigned community, and the color of an edge is then defined by the colors of it origin and destination. Additionally, the white lines are the subway tunnels in the study area.}
   \label{fig:polycentric}
\end{figure}

With prior knowledge of Beijing, we further categorize those clusters into four distinct types as ``commercial-dominant", ``transport-dominant", ``residential-dominant" and ``leisure-dominant". In more details, clusters $C1$, $C2$ and $C3$ are commercial dominant, which well match the three major commercial cores (named as ``Zhong Guan Cun", ``Xi Dan" and ``Guo Mao") of Beijing; clusters $C4$, $C5$, $C6$ and $C11$ are transport-dominant and coincide the four transport hubs (named as ``Xi Zhi Men", ``Beijing-West Railway Station", ``Beijing-South Railway Station" and ``Dong Zhi Men") within the city; clusters $C7$, $C8$, $C9$, $C12$ and $C14$ are residential-dominants, which are generally located at the periphery of the study area; and cluster $C10$ (named as ``San Li Tun") is the largest bar area in Beijing and thus leisure-dominant. Besides, this analysis can also explain the residuals in Figure 1(c), at least partially. If two TAZs with large attractions are close, the gravity model will predict a large flow between them. However, these two nodes also have a high probability belong to a same cluster, indicating a competing role with each other. In other words, it will result in fewer interactions between two close TAZs in reality.

Based on the spatial distributions of different types of clusters and how they are interconnected with each other in terms of taxi trips, the study area can be split into three large centers, unraveling the polycentric structure of the study area. The first center covers clusters $C1$, $C12$ and $C3$; The second center contains clusters $C3$, $C4$, $C5$ and $C6$; The third center involves clusters $C2$, $C10$ and $C11$. This polycentric structure is consistent with the division of the functional areas of Beijing, suggesting the capability of identifying urban structure from taxi OD datasets.

\subsection{Link Semantics}

To further understand how TAZs are interconnected, we also exploit the temporal characteristics of inter-TAZ tax flows. As mentioned in Section III, link signatures $S_{ij}^{res}$ are automatically assigned into four groups (see Figure 4). Obviously, these four groups have distinct temporal characteristics of interaction intensity and imply how TAZs are interconnected with each other along with time .

\begin{figure*}[htbp]
        \centering
        \begin{subfigure}[b]{0.9\textwidth}
                \includegraphics[trim={6.5cm 0 5.5cm 0}, clip, width=0.2\textwidth, height=1.8in]{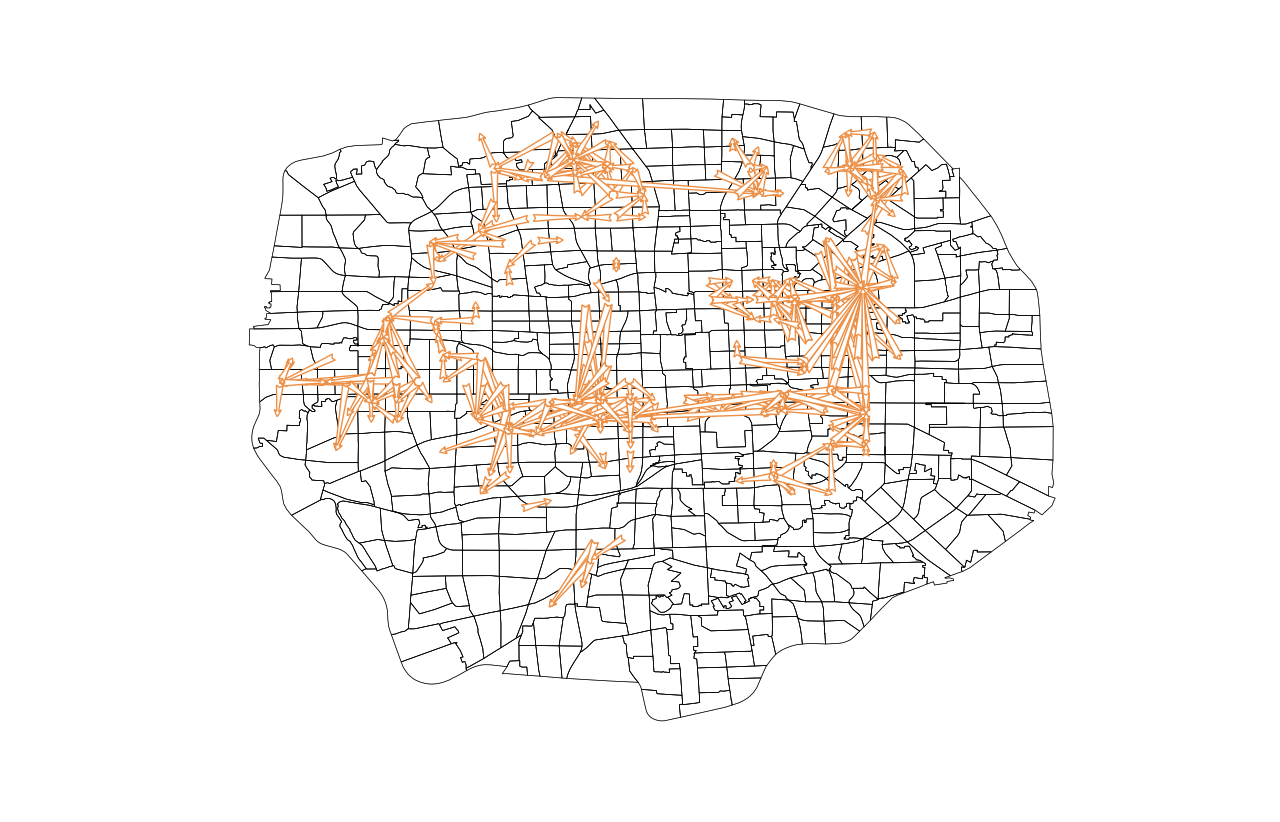}
                \includegraphics[trim={6.5cm 0 5.5cm 0}, clip, width=0.2\textwidth, height=1.8in]{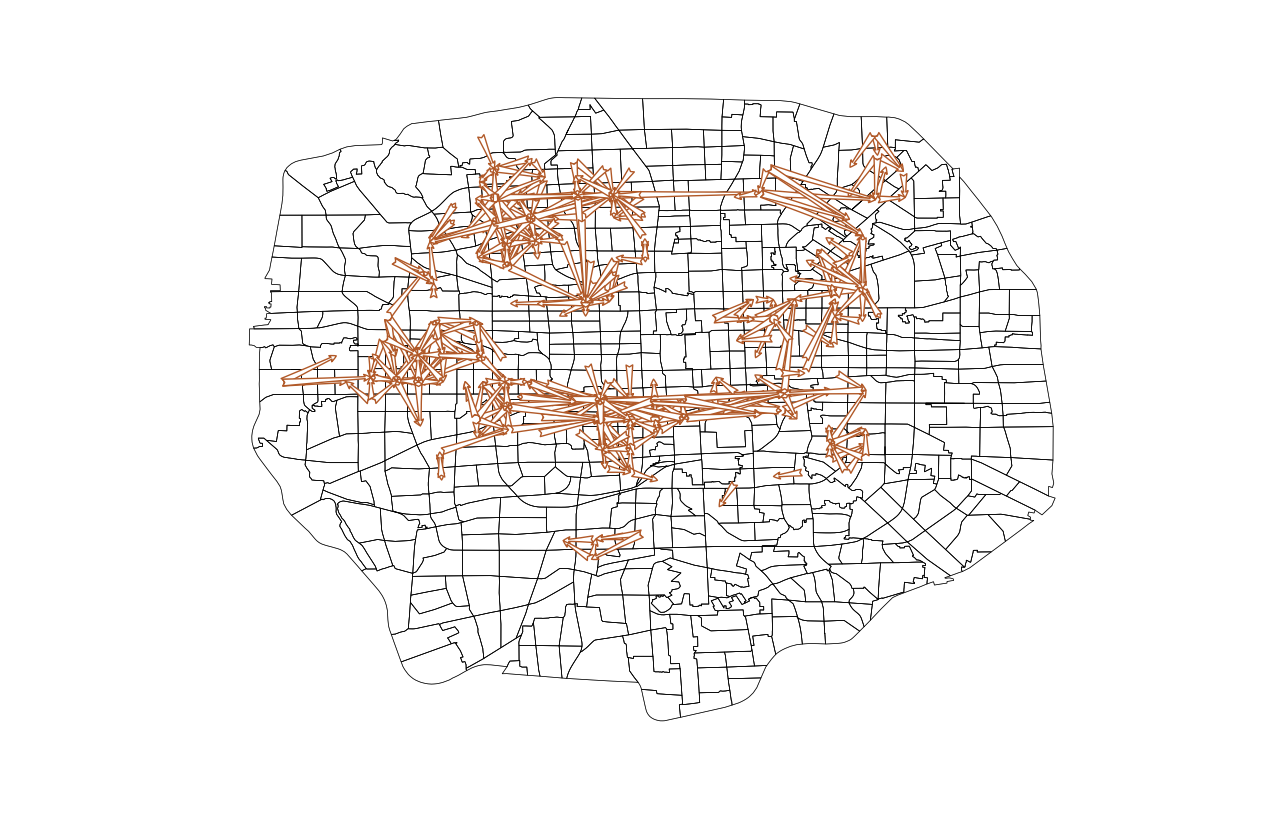}
                \includegraphics[trim={6.5cm 0 5.5cm 0}, clip, width=0.2\textwidth, height=1.8in]{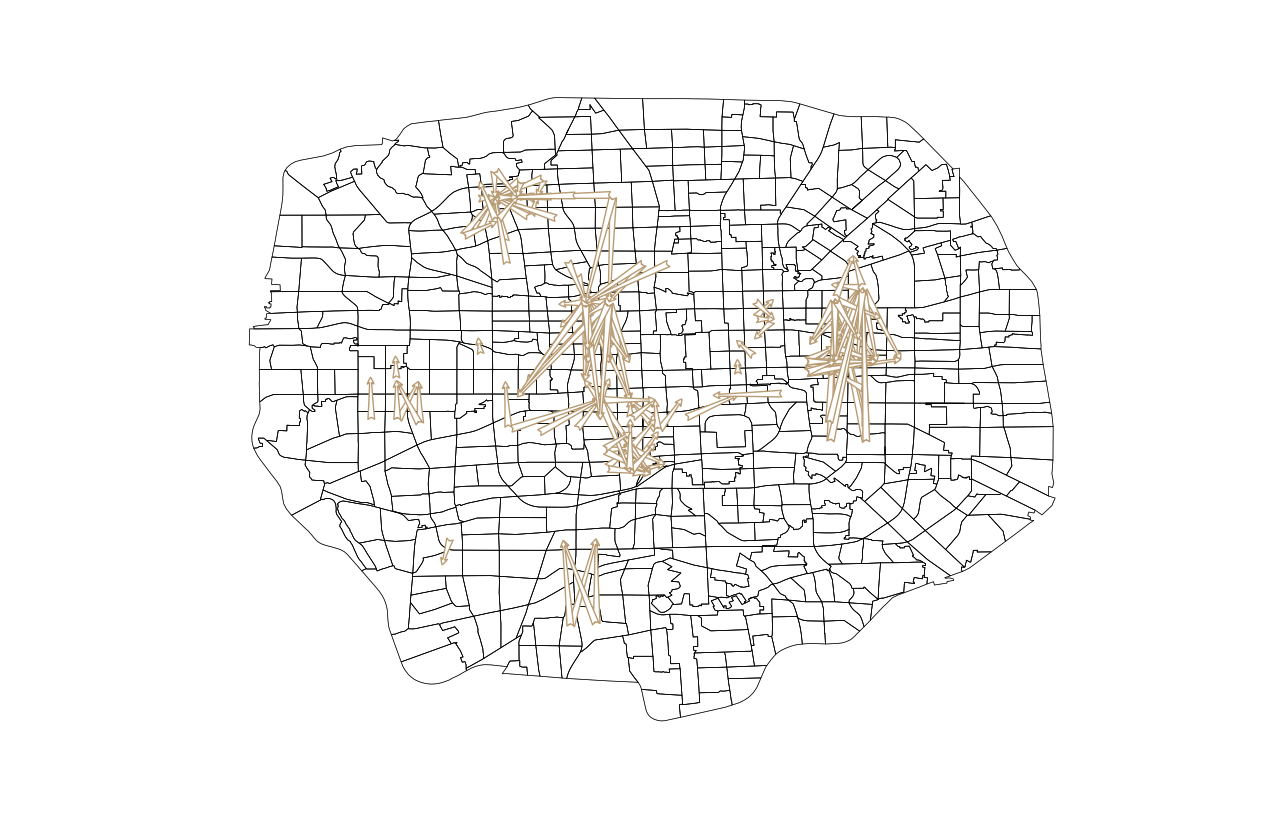}
                \includegraphics[trim={6.5cm 0 5.5cm 0}, clip, width=0.2\textwidth, height=1.8in]{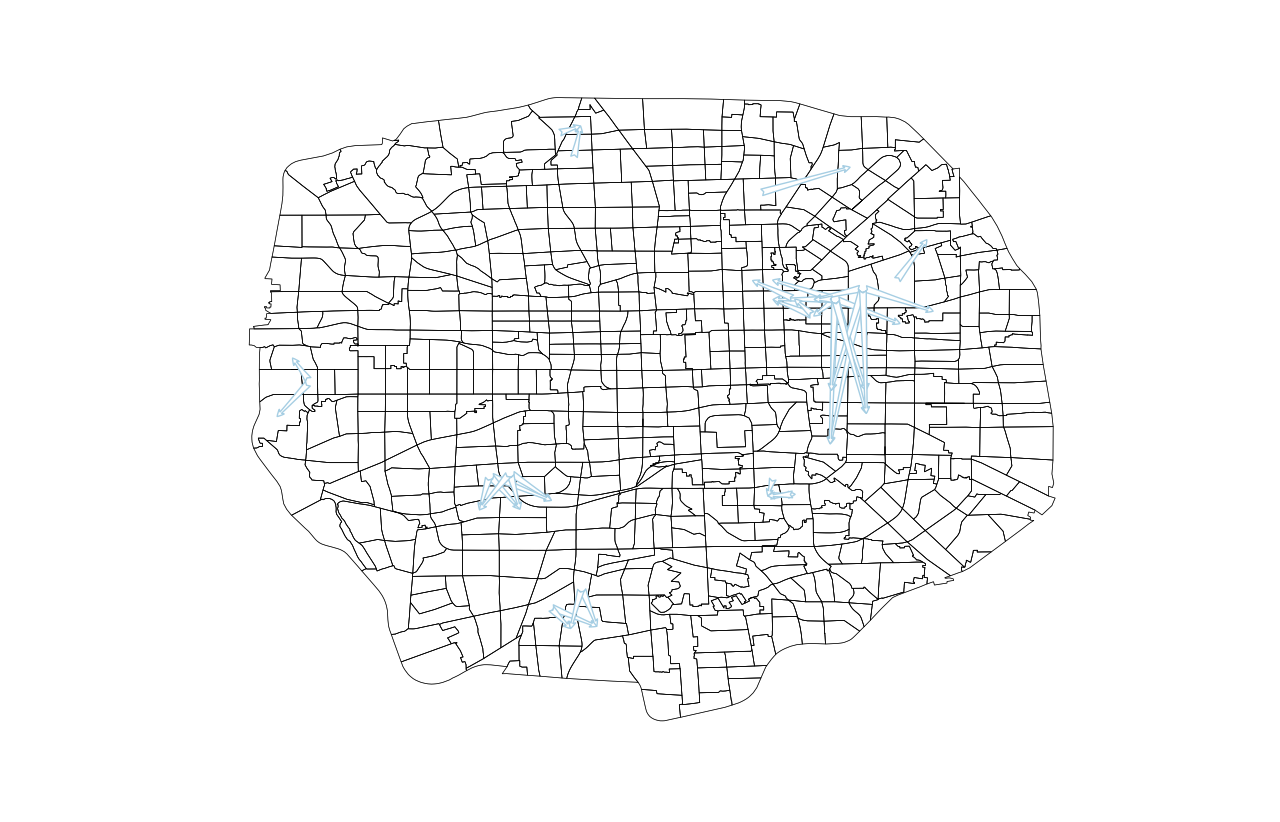}
                \centerline{\includegraphics[width=0.765\textwidth]{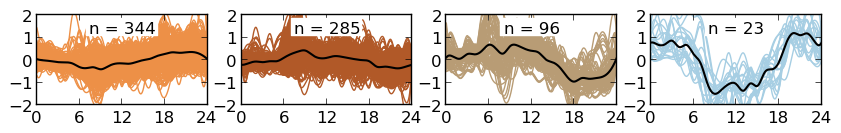}}                
                \caption{Weekday}
                \label{fig:gull}
        \end{subfigure}%
        ~\\ 
        \begin{subfigure}[b]{0.9\textwidth}
                \includegraphics[trim={6.5cm 0 5.5cm 0}, clip, width=0.2\textwidth, height=1.8in]{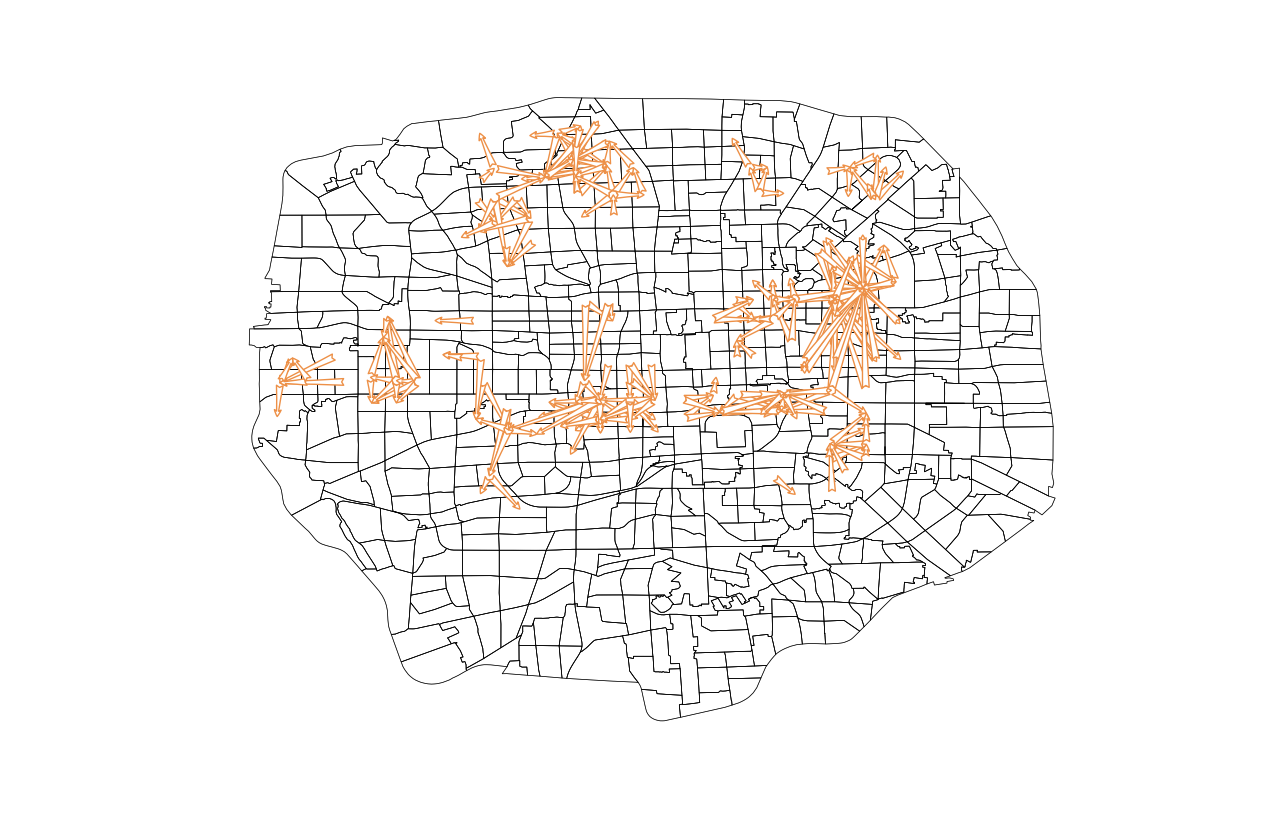}
                \includegraphics[trim={6.5cm 0 5.5cm 0}, clip, width=0.2\textwidth, height=1.8in]{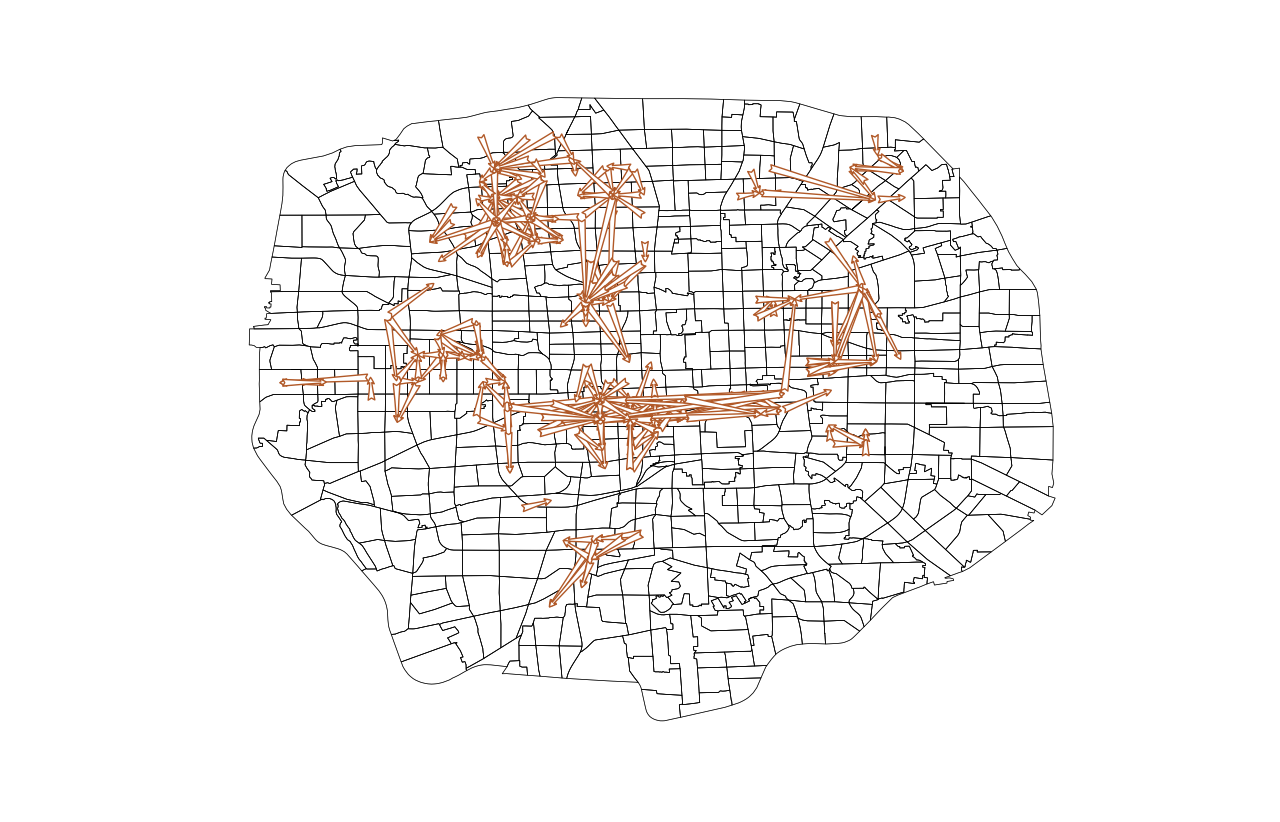}
                \includegraphics[trim={6.5cm 0 5.5cm 0}, clip, width=0.2\textwidth, height=1.8in]{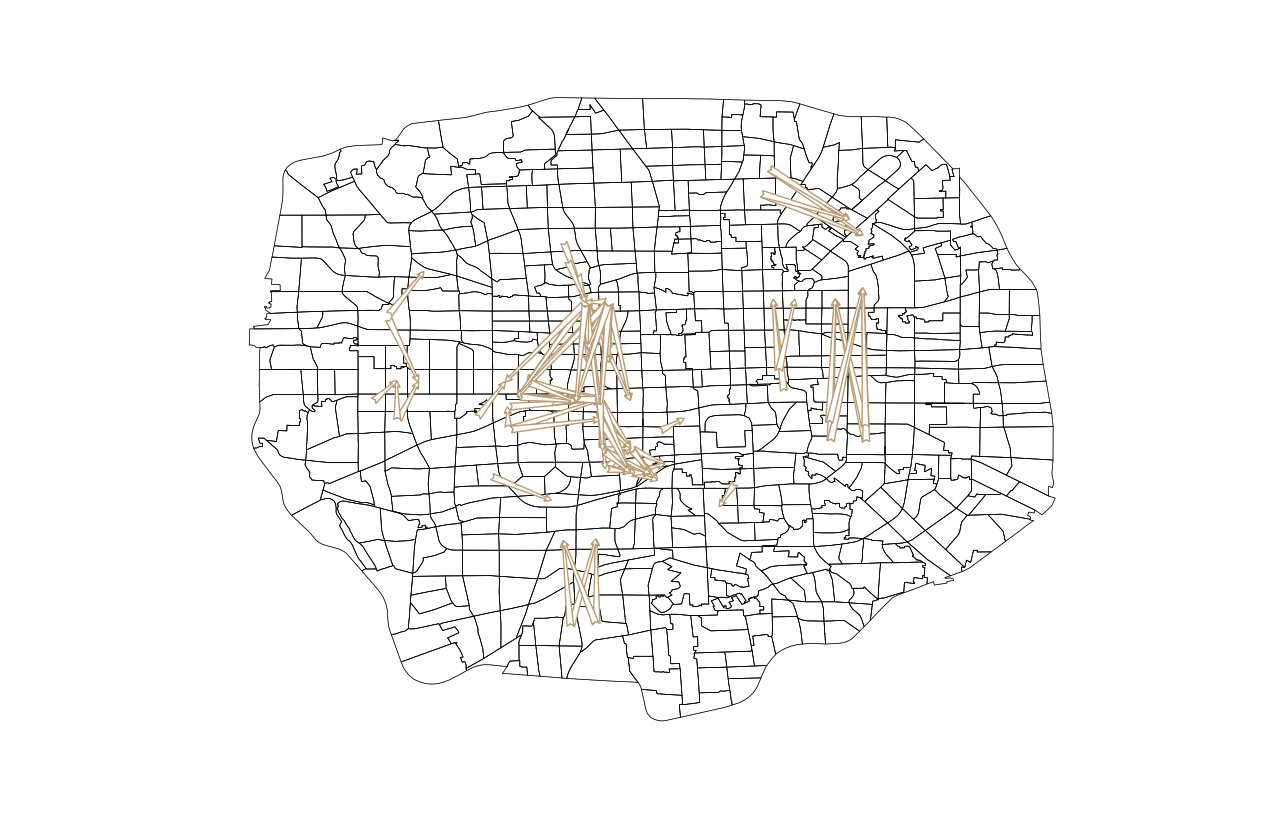}
                \includegraphics[trim={6.5cm 0 5.5cm 0}, clip, width=0.2\textwidth, height=1.8in]{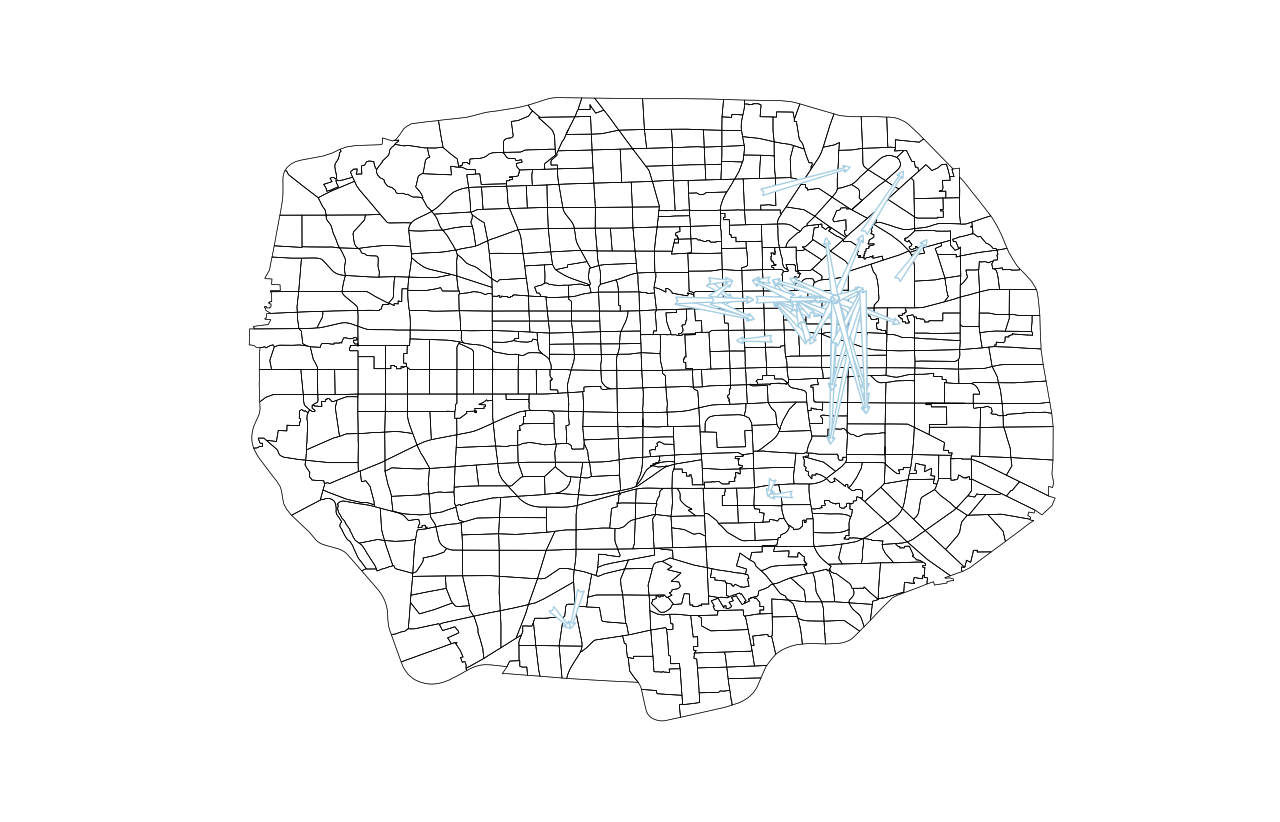}
                \centerline{\includegraphics[width=0.765\textwidth]{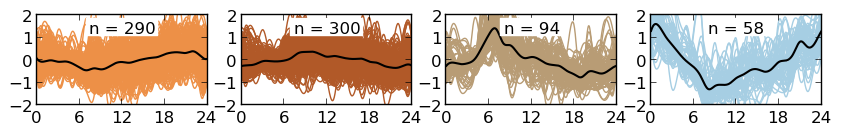}}       
                \caption{Weekend}
                \label{fig:mouse}
        \end{subfigure}
        \caption{Semantics of inter-TAZ taxi flows. The link signatures $S_{ij}^{res}$ are clustered into 4 groups by the X-means algorithm. In the (bottom) plot, $n$ are the number of links assigned to each cluster. These links are also mapped on the TAZ map (top) to illustrate their spatial distribution pattern. We interpret: group 1 (orange) as flows from workplace to home, having a peak value during the night commuting period; group 2 (brown) and group 3 (wood) as flows from home to workplace, with a peak value during the morning commuting period. Particularly, links in group 3 are long distance commuting flows; and group 4 (cyan) as flows from entertainment area to home with large traffic at night. This figure also reveals the differences between the semantics of links in weekdays and weekends.}\label{fig:semantic}
\end{figure*}

Links in the first group (orange) have larger weights after working hours than during morning and working hours. It means passengers prefer to take taxicabs during the night commuting period on these directed routes (or paths). On the contrary, links in the second (brown) and the third (wood) clusters has the opposite patterns compared with the first cluster. In other words, passengers prefer to take taxicabs during the morning commuting period on these directed routes. Moreover, links with most significant patterns are in the last cluster (cyan). There is a higher probability of picking up passengers at night-time than day-time on these routes compared with links in other clusters. Note that it not necessarily means there are more passengers at night-time than day-time in that $R_{ij}^{t}$ of signature $S_{ij}^{res}$ is compared with the average weight of all links at time $t$ (refer to Section III for more details).       

We thus interpret: (1) group 1 as flows from workplace to home, having a peak during the night commuting period. Spatially, these links are generally directed from (sub-)centers to their surrounding areas; (2) group 2 and group 3 as flows from home to workplace, with relatively large weight during the morning commuting period. Furthermore, the peak in group 3 is more significant than in group 2 in that links in group 3 represent long-distance commuting flows; and (3) group 4 as flows from entertainment areas to residential areas and transport hubs, with relative large traffic at night and concentrating at the largest bar area (``San Li Tun") of Beijing. Since the Gravity Model take no consideration on link semantics, it fails to predict flows with specific meanings.  

Another interesting finding is the differences between the semantics of links in weekdays and weekends. As discussed above, group 3 captures the morning commuting routes from home to workplace in general sense. The destinations of these flows are concentrated at commercial centers and transport hubs (please refer to Figure 3). However, the flows directing to cluster $C1$ disappear in weekends due to this area changes its social function from workplace in weekdays to leisure area in weekends. Additionally, the number of links in group 4 increases remarkably in weekends compared with weekdays in that people have more flexible time budgets to conduct leisure activities.


\section{Conclusion}
This article analyzed the spatial interactions between 652 traffic analysis zones in Beijing by taxi trips and found that: (1) the inter-TAZ network of Beijing has a clear gravitational structure. The network are well governed by the Gravity Model $G_{ij} = \lambda p_{i}p_{j} / d_{ij}$, where $p_{i}$, $p_{j}$ are degrees of TAZ $i$, $j$ and $d_{ij}$ the distance between them. This highly simplified model provides a most intuitive way to understand and predict inter-TAZ taxi traffic; (2) the inter-TAZ network is also polycentric organized. There are 3 large (sub-)centers within the study area, each of which has a significant impact on its surrounding (sub-)regions. We also find that taxi ODs generally concentrate at certain commercial centers, transport hubs and residential areas, resulting in cohesive clusters of TAZs in space and distinct functional (sub-)areas; (3) the inter-TAZ taxi flows have significant semantics. There are four typical inter-TAZ flows as ``home $\rightarrow$ workplace (or transport hub)", ``workplace $\rightarrow$ home (or transport hub) (short-distance)", ``workplace $\rightarrow$ home (or transport hub) (long-distance) " and ``leisure $\rightarrow$ home (transport hub)'' which have distinct temporal patterns of interaction intensity. To summarize, these three structural properties well reveal how the TAZs are organized and interconnected with each other in Beijing.


\section*{Acknowledgment}
The authors thank to Dr Ying Long for providing the geographical information of the study area and Professor Yanwei Chai for discussion.


\end{document}